\documentclass[
twocolumn,
superscriptaddress,
amsmath,amssymb,
aps,
]{revtex4-2}

\usepackage{dcolumn}
\usepackage{bm}
\usepackage{hyperref}
\usepackage{IEEEtrantools}

\usepackage[utf8]{inputenc} 
\usepackage[T1]{fontenc}    
\usepackage{hyperref}       
\usepackage{url}            
\usepackage{booktabs}       
\usepackage{amsfonts}       
\usepackage{nicefrac}       
\usepackage{microtype}      

\usepackage{graphicx}
\graphicspath{{figures/}}
\usepackage{physics}
\usepackage{subcaption}
\usepackage{wrapfig}

\usepackage{tikz}
\usetikzlibrary{quantikz}
\usepackage{bm} 
\usetikzlibrary{shapes,arrows}
\usetikzlibrary{positioning}
\usetikzlibrary{shadows}

\def \dd  {{\rm d}}

\newcommand{\figref}[1]{FIG.~\ref{#1}}

\usepackage[en-GB]{datetime2}

\usepackage{nonfloat}

\begin{document}
	
	\title{Quantum Generative Adversarial Networks in a Continuous-Variable Architecture to Simulate High Energy Physics Detectors}
	
	\author{Su Yeon Chang}
	\affiliation{OpenLab, CERN, Geneva, Switzerland}
	\affiliation{Institute of Physics, \'{E}cole Polytechnique F\'{e}d\'{e}rale de Lausanne (EPFL), Lausanne, Switzerland}
	\author{Sofia Vallecorsa}
	\affiliation{OpenLab, CERN, Geneva, Switzerland}
	\author{Elías F. Combarro}
	\affiliation{University of Oviedo, Oviedo, Spain }
	\author{Federico Carminati}
	\affiliation{OpenLab, CERN, Geneva, Switzerland}
	
	\date{\today}
	
	\begin{abstract}
	Deep Neural Networks (DNNs) come into the limelight in High Energy Physics (HEP) in order to manipulate the increasing amount of data encountered in the next generation of accelerators. Recently, the HEP community has suggested Generative Adversarial Networks (GANs) to replace traditional time-consuming \textit{Geant4} simulations based on the Monte Carlo method.
	In parallel with advances in deep learning, intriguing studies have been conducted in the last decade on quantum computing, including the Quantum GAN model suggested by IBM. However, this model is limited in learning a probability distribution over discrete variables, while we initially aim to reproduce a distribution over continuous variables in HEP.  
	
	We introduce and analyze a new prototype of quantum GAN (qGAN) employed in continuous-variable (CV) quantum computing, which encodes quantum information in a continuous physical observable. Two CV qGAN models with a quantum and a classical discriminator have been tested to reproduce calorimeter outputs in a reduced size, and their advantages and limitations  are discussed. 
	\end{abstract}
	
	\maketitle
	
	\section{Introduction}
	
	Modern society confronts an overwhelming amount of data, which should be treated with high precision and limited resources. Especially, in High Energy Physics (HEP), next High Luminosity Large Hadron Collider (HL-LHC) phase will collect and analyze an increasing amount of data with complex physics and small statistical error and it is therefore highly demanding in terms of computational resources. Traditional Monte Carlo based simulation is however very time-consuming, thus new approaches using Deep Neural Networks (DNNs) have been proposed to replace it.    
	
	Generative Adversarial Networks (GANs) \cite{GAN} arise as one of the solutions for fast simulation. Based on its two neural networks, generator and discriminator, trained alternatively, the GAN model has been widely explored thanks to its performance to generate images with complex structures at much higher speed.  In HEP, the variations of GAN, such as CaloGAN \cite{CaloGAN}, have achieved to show similar performance to that of full Monte-Carlo based simulation, but with less amount of time taken for the computation. Meanwhile, quantum computing has emerged as another important pillar in modern studies attracting the attention of many researchers due to its potential to execute some tasks with an exponentially reduced amount of resources in time and space compared to a classical processor \cite{Google}. 
	
	Advances in both deep learning and quantum computing have made it possible to extend the GAN paradigm to the quantum world, with the introduction of Quantum GAN (qGAN) models \cite{QGAN, QGAN_qiskit}. Unfortunately, the most well-known qGAN model proposed by IBM \cite{QGAN_qiskit}, with a classical discriminator and a quantum generator in a qubit architecture, is limited to reproducing the probability distribution over discrete variables, while the calorimeters outputs that we would like to treat are continuous. 
	
	Continuous-variable (CV) quantum computation offers an insight to solve the problem. Unlike the well-known \textit{qubit} architecture, which uses discrete fundamental information-carrying units, CV quantum computing uses \textit{qumodes} which are continuous by their nature \cite{CV_model}. Taking a complementary position to the qubit architecture,  the CV architecture has shown its efficiency in a number of papers that studies its applications, including CV neural networks (CVNNs) \cite{CVNN}.    
	
	Our work introduces a new prototype of qGAN, implemented in CV architecture. The model has been tested to simulate continuous detector outputs, with a reduced size, and its results are analyzed and discussed in this paper.

	\section{Continuous-variable (CV) quantum computing}
	\label{sec:CV}
	
	Continuous-variable (CV) architecture is a paradigm of quantum computing, where the information is encoded in a continuous physical observable, for instance the strength of the electromagnetic field. In this model, the fundamental information-carrying unit is represented by a \textit{qumode}, which is a quantum state of bosonic modes in the quantized electromagnetic field. The quantum state of $N$ qumodes is often expressed as a superposition of position basis, $\{\ket{\mathbf{{x}}}\}_{\mathbf{x} \in \mathbb{R}^N}$ or Fock basis, $\{\ket{n}\}_{i \in \mathbb{N}}$  :  
	
	\begin{equation}
	\ket{\psi} = \int_{\mathbb{R}^N} \psi(\mathbf{x}) \ket{\mathbf{x}} \dd \mathbf{x} = \sum_{i = 0}^\infty \bra{n}\ket{\psi} \ket{n}.
	\end{equation}

	\begin{figure}[h]
		\begin{center}
			\begin{quantikz}
				\lstick{} & [-0.2cm] \gate[wires = 4]{\mathcal{U}_1(\bm{\theta}_1,\bm{\phi}_1)} & [-0.2cm] \gate{S(z_1)} & [-0.2cm] \gate[wires = 4]{\mathcal{U}_2(\bm{\theta}_2,\bm{\phi}_2)} & [-0.2cm]  \gate{D(\alpha_1)}&  [-0.2cm] \gate{\Phi(\phi_1)}& [-0.2cm]\qw \\ [-0.35cm]
				\lstick{} &  & [-0.2cm] \gate{S(z_2)}&  & [-0.2cm]\gate{D(\alpha_2)}& [-0.2cm]\gate{\Phi(\phi_2)} & [-0.2cm]\qw \\[-0.35cm]
				\lstick{}&    & [-0.2cm] \gate{S(z_3)}&  & [-0.2cm]\gate{D(\alpha_3)}& [-0.2cm]\gate{\Phi(\phi_3)}&[-0.2cm]\qw \\[-0.35cm]
				\lstick{}&   & [-0.2cm] \gate{S(z_4)}&  &[-0.2cm]\gate{D(\alpha_4)}&[-0.2cm] \gate{\Phi(\phi_4)}& [-0.2cm]\qw 
			\end{quantikz}
		\end{center}
		\caption{\textit{Structure of a CV neural network layer.} \cite{CVNN} } 
		\label{fig:CVNN}
	\end{figure}
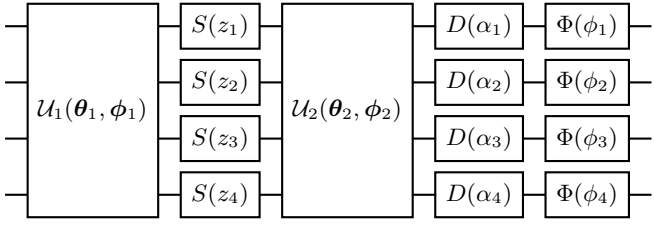
	
	By combining a set of common CV gates, such as displacement $D(\alpha)$, rotation $R(\phi)$, squeezing $S(z)$, beamsplitters $BS(\theta, \phi)$, etc, it is possible to build CV Neural Networks (CVNN), parameterized quantum circuits which perform exactly the same mathematical transformation as a fully connected layer.

	Consider that a $N$-dimensional vector $\mathbf{x}$ is represented by an eigenstate $\ket{\mathbf{x}}$ of the position operator $\hat{x}$ with
	$\ket{\mathbf{x}} = \bigotimes_{i=1}^N \ket{x_i} =  D(\mathbf{x})\ket{0}^{\otimes N}$ where $\ket{0}$ is the vacuum state.

		Then, the sequence of CV gates shown on \figref{fig:CVNN} performs the following transformation $\mathcal{L}$ which is equivalent to a single classical  fully connected layer 
	\begin{equation}
	\mathcal{L}\ket{\mathbf{x}} =  \Phi \circ  \mathcal{D} \circ \mathcal{U}_2 \circ \mathcal{S} \circ \mathcal{U}_1 \ket{\mathbf{x}} \propto \ket{\phi(W\mathbf{x} + \mathbf{b})}, 
	\end{equation}
	where $\mathcal{U}$ is an interferometer \cite{Interferometer1}, $W$ the weight matrix, $\mathbf{b}$ the bias and $\Phi$ the non-linear activation function. 
	More detailed explanations on CVNN can be found in Ref.\cite{CVNN}

	\section{CV qGAN}

	This section proposes two different prototypes of continuous-variable Quantum GAN (CV qGAN) models: one with both quantum generator and discriminator (\textit{Fully quantum} model), and the second with quantum generator and classical discriminator (\textit{Hybrid} model). \figref{fig:CV_qGAN} summarizes the structure of the two models.

	In both cases, $N$ qumodes quantum generators are initialized with the first qumode displaced by a random noise $z \in \mathcal{N}(0,1)$, while keeping all the other qumodes to be vacuum : 
	\begin{equation}
	\ket{initial} = \ket{z}\otimes\ket{0}^{\otimes N-1} = D(z)\ket{0}\otimes\ket{0}^{\otimes N-1}.
	\end{equation}
	It then applies CVNN layers shown on \figref{fig:CVNN} to the initial state to transform it into the targeted state. The difference between the two prototypes comes from the measurement at the end of the generator.  
	In fully quantum case, the quantum generator is directly connected to the quantum discriminator, without any measurement between them. The expectation value of the position operator at qumode $N$, $\langle x \rangle$ is measured at the end of the discriminator while discarding all the other qumodes. Another classical sigmoid function is then applied to return a predicted label, $\ell \in [0,1]$, indicating the authenticity of the sample.
	
	On the other hand, the hybrid model constructs a fake image by measuring the position expectation values at all $N$ qumodes, $\langle \mathbf{x} \rangle$, at the end of the generator. This fake image is passed to the classical discriminator, to return a prediction used to update the generator parameters. 
	
\begin{figure}[h]
		\begin{center}
		\begin{subfigure}{0.235\textwidth}
			\includegraphics[width = \textwidth]{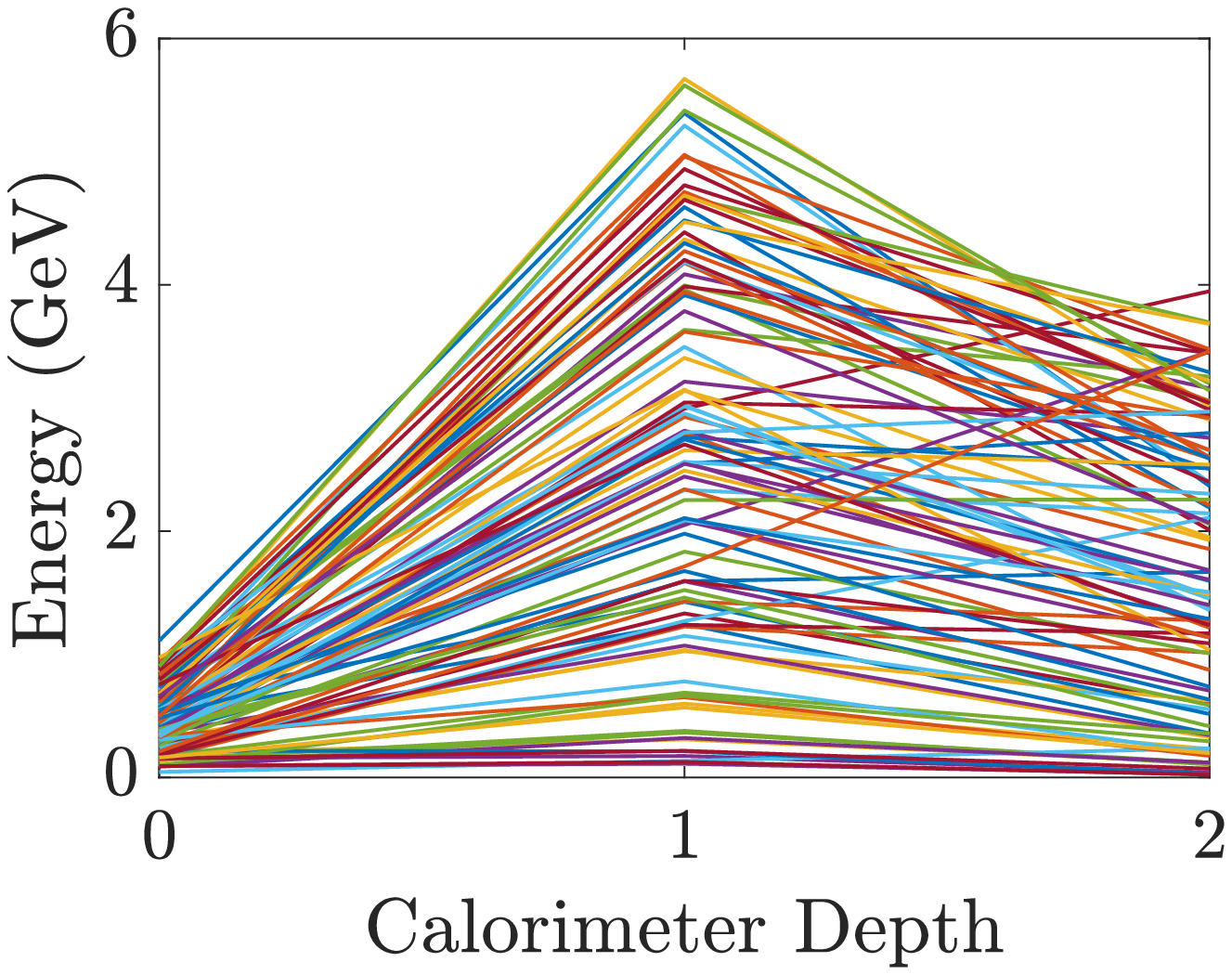}
			\caption{\em Real images. }
			\label{fig:real_data_3pixels}
		\end{subfigure}
		\begin{subfigure}{0.235\textwidth}
			\includegraphics[width = \textwidth]{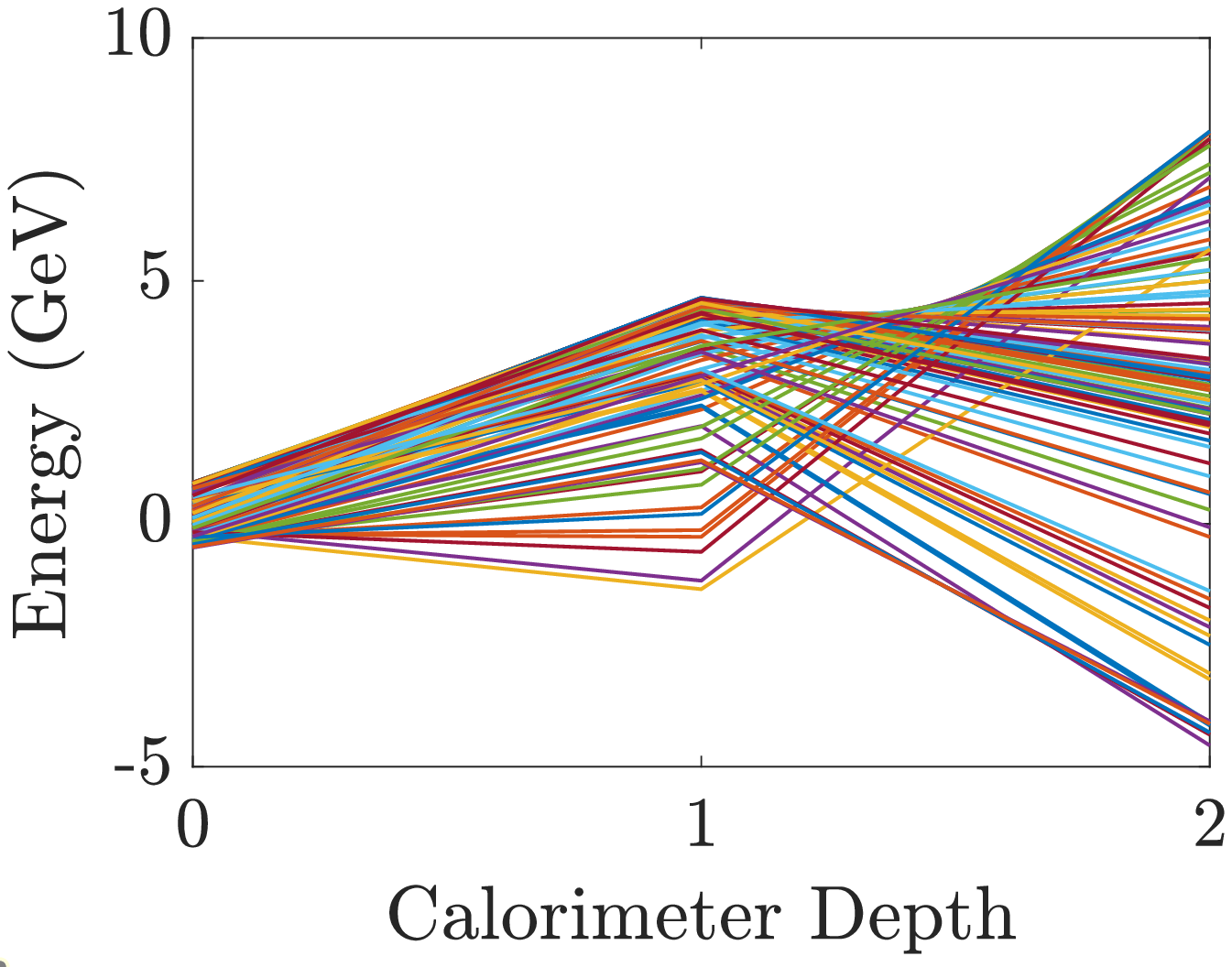}
			\caption{\em Fully quantum, $d_g = 5$}
			\label{fig:CVquantum_images}
		\end{subfigure}
		\begin{subfigure}{0.235\textwidth}
			\includegraphics[width = \textwidth]{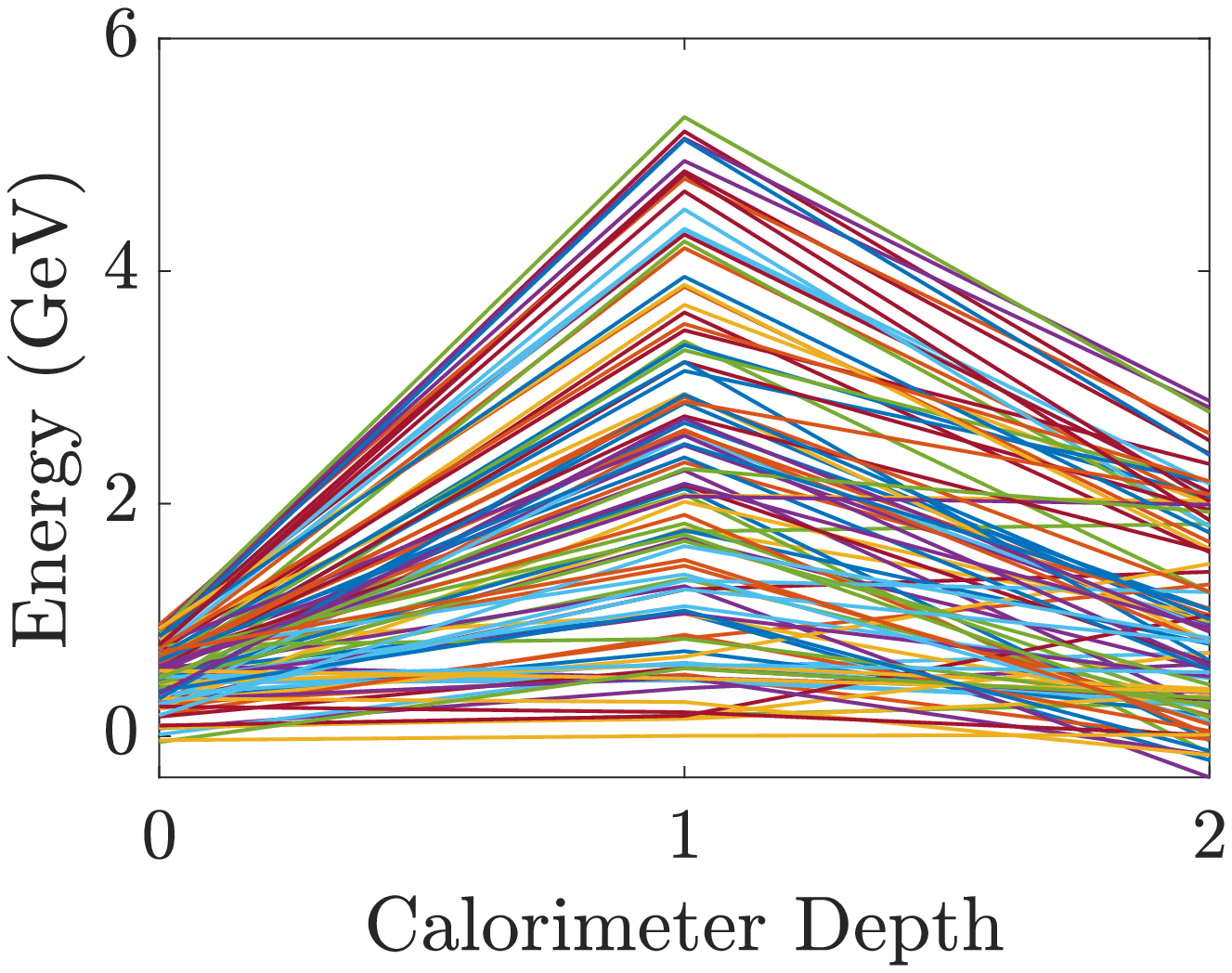}
			\caption{\em Hybrid, $d_g = 5$}
			\label{fig:CVhybrid_images2}
		\end{subfigure}
		\begin{subfigure}{0.235\textwidth}
			\centering
			\includegraphics[width = \textwidth]{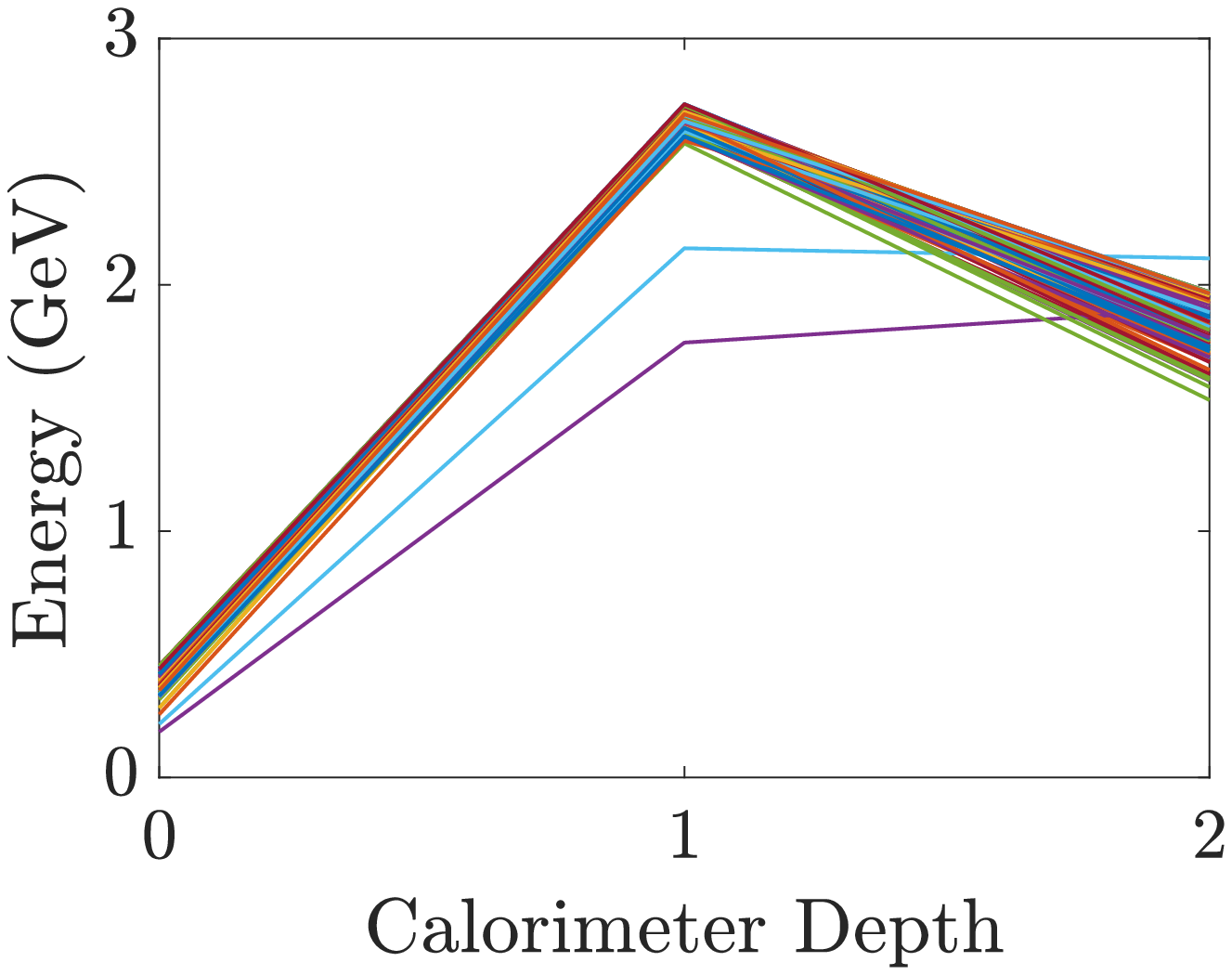}
			\caption{\em Hybrid, $d_g = 3$}
			\label{fig:CVhybrid_images}
		\end{subfigure}
		\end{center}
		\caption{\em Comparison of the real training samples (a) with the images generated by the fully quantum CV qGAN with $d_g = 5$ (b) and the hybrid  CV qGAN, with $d_g = 5$ (c) and $d_g = 3$ (d). }
		\label{fig:CV_images}
	\end{figure}

	This CV qGAN has been tested only for three qumodes, i.e., $n = 3$, due to the limitation in simulation computing time and memory. The original calorimeter outputs of size $25\times25$ are averaged over the longitudinal direction, and then binned into 3 pixels, as shown on \figref{fig:real_data_3pixels}.


	The model is implemented in \textit{Strawberryfields} 
	\cite{Strawberryfields, Strawberryfield_site} for CV quantum computing and \textit{Pennylane} \cite{Pennylane, Pennylane_site} for automatic quantum gradient descent.

    \onecolumngrid

	\tikzstyle{line} = [draw, -latex']
	\begin{figure}[!b]
		\begin{subfigure}{\textwidth}
			\begin{center}
				\begin{quantikz}
					\lstick{$\ket{0}^{\otimes {N-1}}$}& \qwbundle{} & \gate[wires = 2,,disable auto height, style={text width=8em, fill=green!20, rounded corners}]{\text{\sc \bf Generator}}& \qw& \qw & \qw \slice{{\color{red} No Measurement }} & \qw & \qw&\gate[wires = 2,,disable auto height, style={text width=8em, fill=blue!20, rounded corners}]{\text{\sc \bf Discriminator}} & \qw  \rstick{Discard} \\[-0.4cm]
					\lstick{$\ket{z\sim\mathcal{N}(0,1)}_x$}& \qw & & \qw &\qw& \qw &\qw	& \qw & \qw & \qw \rstick{$\langle x_N \rangle~ \xrightarrow{sigmoid} ~ \ell  $}	\end{quantikz}
			\end{center}
			\caption{\em Fully quantum model.}
		\end{subfigure}

		\begin{subfigure}{\textwidth}
			\begin{subfigure}{0.5\textwidth}
				\begin{quantikz}
					\lstick{$\ket{0}^{\otimes {N-1}}$}& \qwbundle{} & \gate[wires = 2, style={text width=8em, fill=green!20, rounded corners}]{\begin{array}{c} \text{\sc \bf Quantum} \\ \text{\sc\bf  Generator} \end{array}}&  \meter{$\ket{\mathbf{x}}$} & \qw \rstick[wires=2]{$\langle \mathbf{x}\rangle$} \\ [-0.6cm]
					\lstick{$\ket{z\sim\mathcal{N}(0,1)}_x$}& \qw & & \meter{$\ket{x}$}& \qw	
				\end{quantikz}
			\end{subfigure}
			\hspace{-2.2cm}
			\begin{subfigure}{0.43\textwidth}
				\begin{tikzpicture}[align=center,node distance = 2cm, auto]
				\node [rectangle, draw, fill=blue!20, 
				text width=8em, text centered, rounded corners, minimum height=1.6cm] (Disc) {\bf Classical \\  \bf Discriminator};
				\node [rectangle,text width = 0.5em, left of = Disc, node distance = 2.5cm] (input) {\phantom{}};
				\node [text width = 1em, right of = Disc, node distance = 2.6cm, minimum height = 1.6cm](PredictedLabel) {$\ell$};
				\path [line] (Disc) -- (PredictedLabel);
				\path [line] (input) -- (Disc);
				\end{tikzpicture}
			\end{subfigure}
			\caption{\em Hybrid model.} 
		\end{subfigure}
		
		\caption{\em Schematic diagrams of continuous-variable quantum GAN models, manipulating classical data embedded in quantum states.}
		\label{fig:CV_qGAN}
	\end{figure}
	
	\newpage
	
	\twocolumngrid

\begin{figure}[h]
		{\centering
		\begin{subfigure}{0.23\textwidth}
			\includegraphics[width = \textwidth]{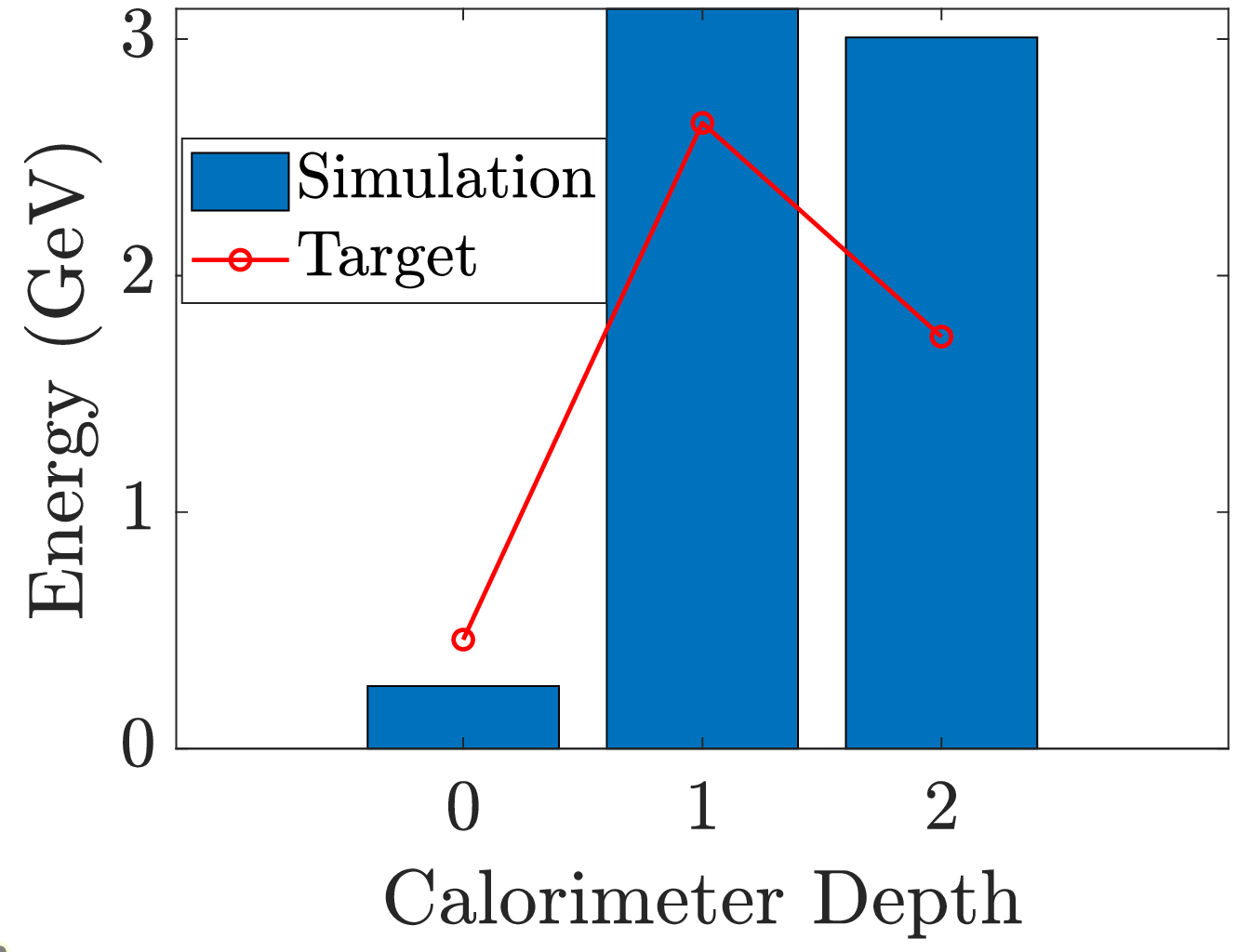}
			\caption{\em}
			\label{fig:mean_quantum}
		\end{subfigure}
		\begin{subfigure}{0.23\textwidth}
			\includegraphics[width = \textwidth]{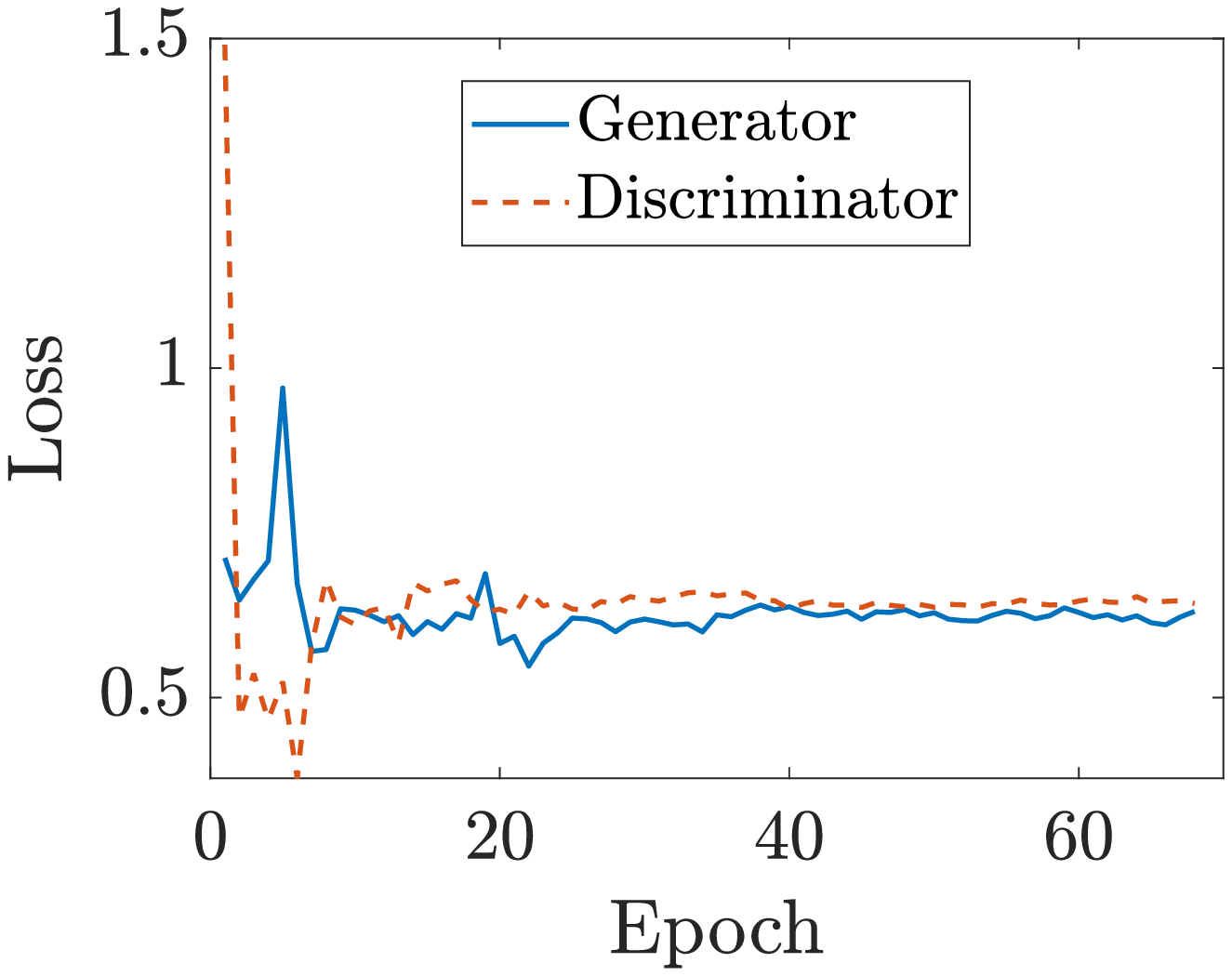}
			\caption{\em}
			\label{fig:loss_quantum}
		\end{subfigure}
		\begin{subfigure}{0.23\textwidth}
			\includegraphics[width = \textwidth]{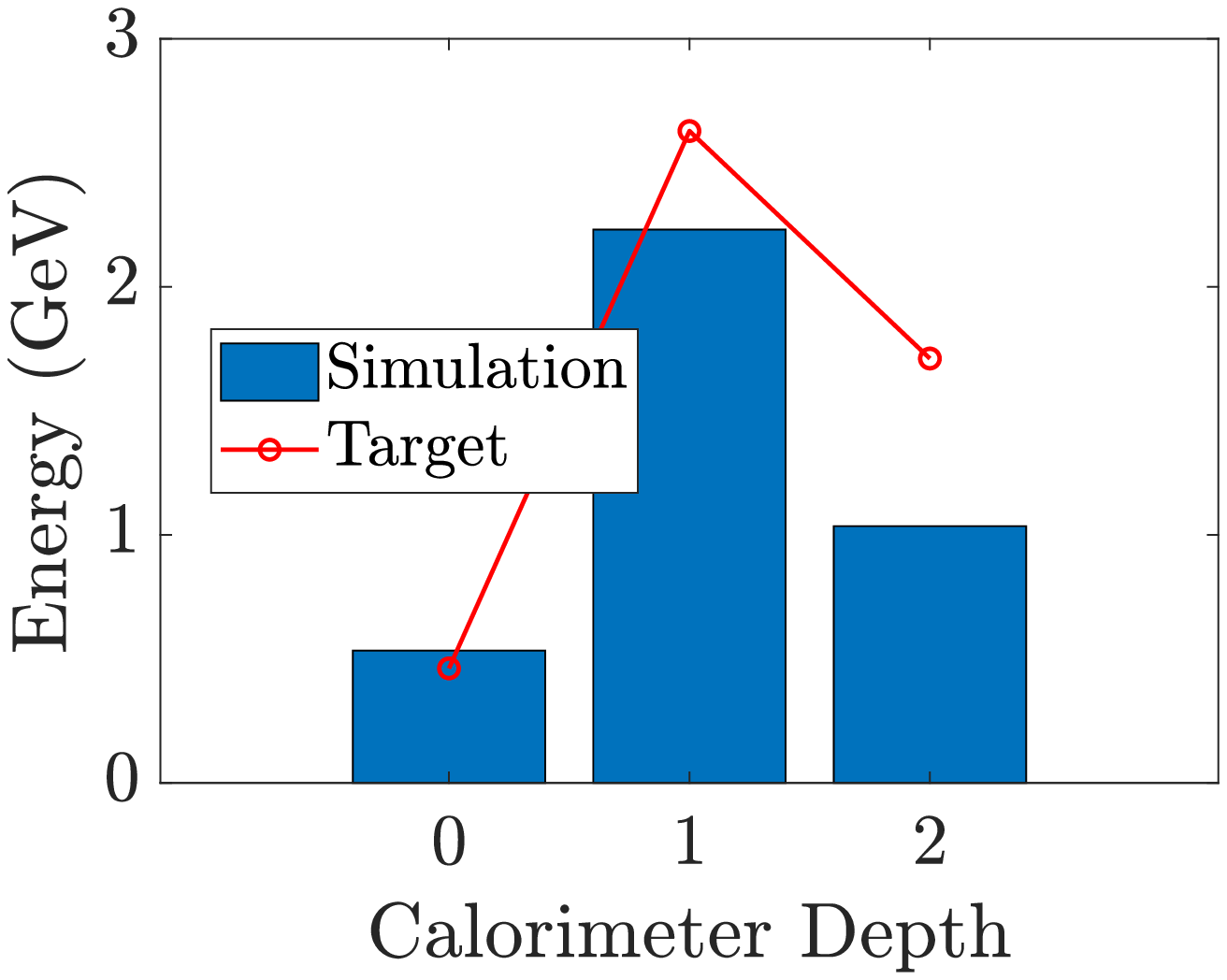}
			\caption{\em}
			\label{fig:mean_hybrid}
		\end{subfigure}
		\begin{subfigure}{0.23\textwidth}
			\includegraphics[width = \textwidth]{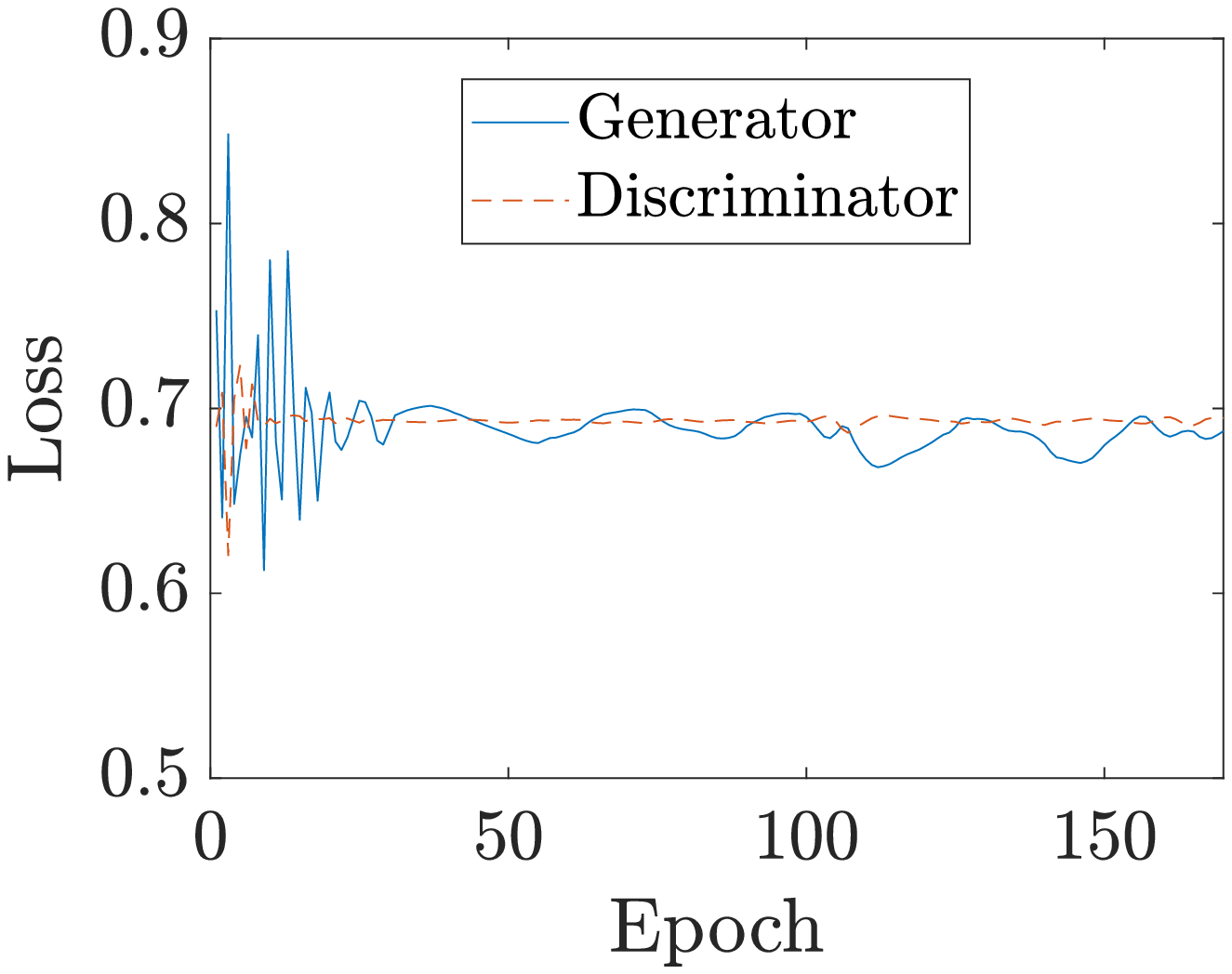}
			\caption{\em}
			\label{fig:loss_hybrid}
		\end{subfigure}
		\begin{subfigure}{0.23\textwidth}
			\includegraphics[width = \textwidth]{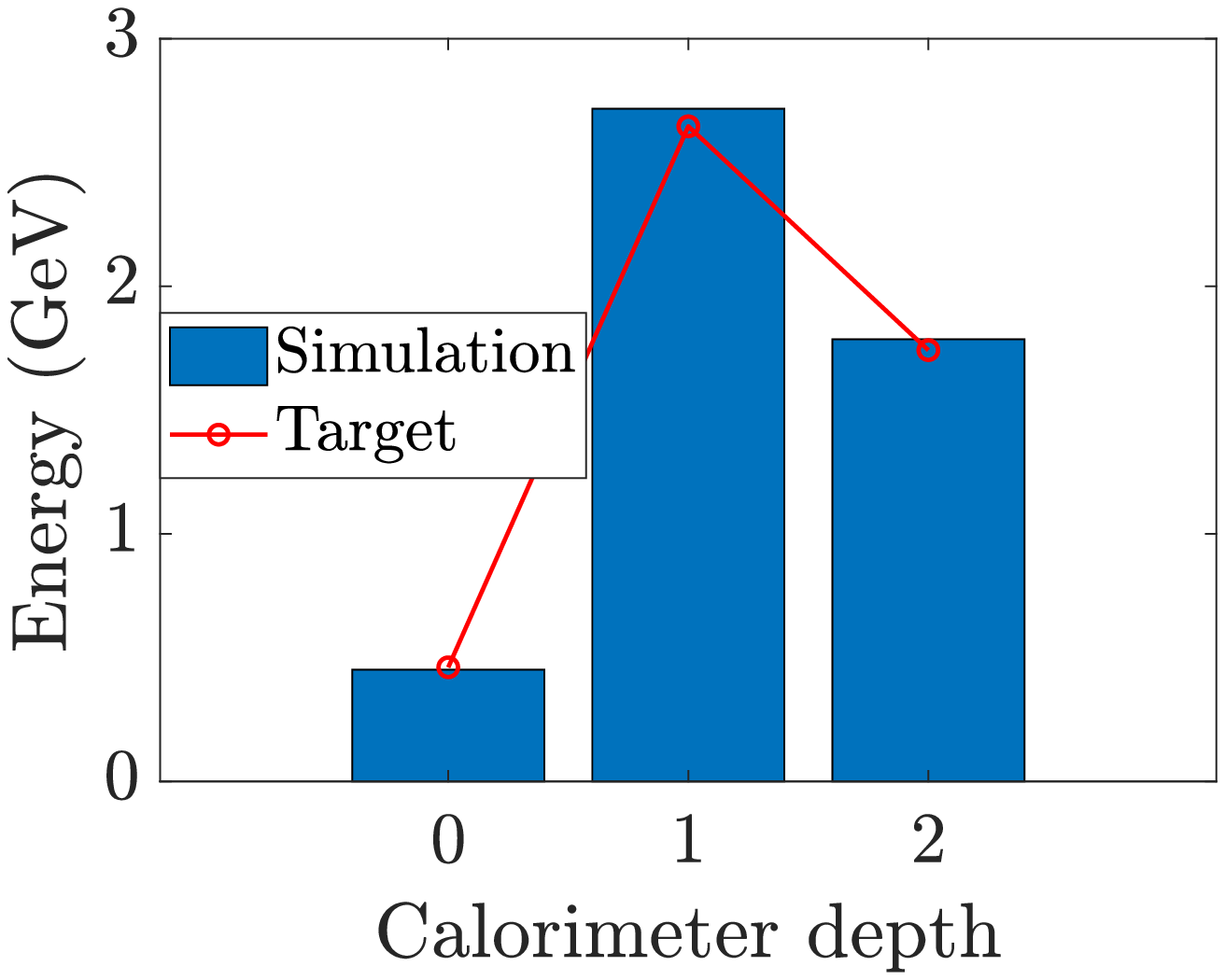}
			\caption{\em}
			\label{fig:CVhybrid_mean}
		\end{subfigure}
		\begin{subfigure}{0.23\textwidth}
			\includegraphics[width = \textwidth]{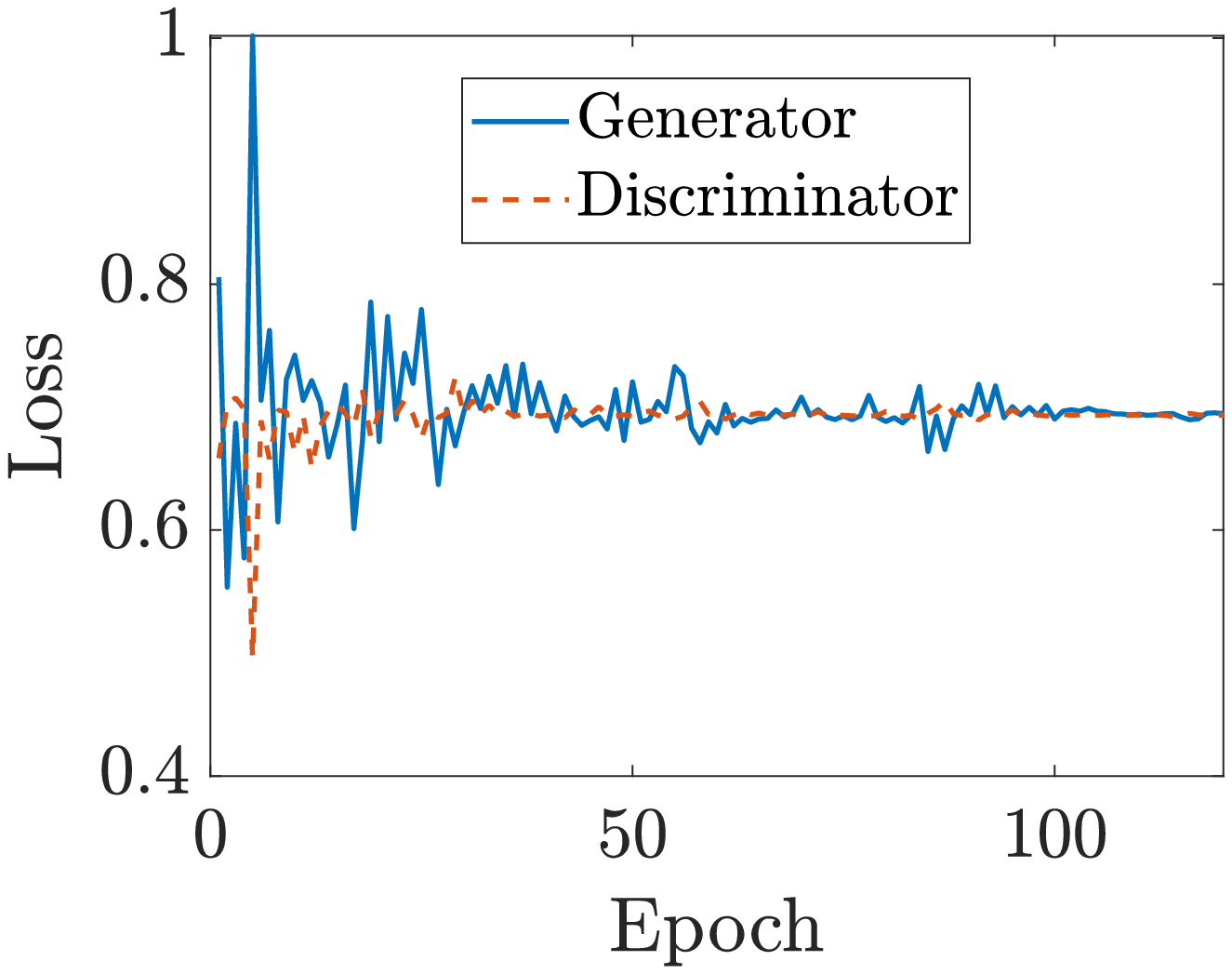}
			\caption{\em}
			\label{fig:CVhybrid_loss}
		\end{subfigure}
		}
		\caption{\em Mean images (a,c,e) and losses (b,d,f) obtained from the CV qGAN model simulations.  Figures (a,b) display the result using the fully quantum model with $d_g = 5$, figures (c,d) the result using the hybrid model, with $d_g = 5$ and figures (e,f) with $d_g = 3$. }
		\label{fig:CVqGAN}
	\end{figure}

	As shown on \figref{fig:CVqGAN}, the generator and the discriminator loss functions converge towards the expected value in all three cases. However, the mean images produced by the fully quantum and hybrid models with $d_g = 5$ differ from the targeted values, showing a limitation in the simulations. Unlike the previous two cases, the result for the hybrid model with $d_g =3$ manifests a convergence of the mean image towards the targeted output. 
	
	The limitation is also revealed throughout the whole set of produced images, shown on \figref{fig:CV_images}. Notably, the images generated by the fully quantum model on \figref{fig:CVquantum_images} have significantly different shapes from the real images shown on \figref{fig:real_data_3pixels}. The most critical problem is that at least half of the images contain negative \textit{energy} values, which are certainly unphysical. The situation is slightly better in the hybrid model with $d_g = 5$ displayed on \figref{fig:CVhybrid_images2}. In spite of the difference in energy levels, the shape of images is similar to the real one, with a peak at $x = 1$. Furthermore, most of the images are above zero except three or four samples. 
	It is also notable that the CV qGAN model can fall into a common GAN failure observed in classical cases. Despite the convergence in mean images for the hybrid model with $d_g = 3$, the individual images shown on \figref{fig:CVhybrid_images} exhibit an obvious mode collapse failure \cite{ModeCollapse} where the generator produces only a small variety of samples.
	
\begin{figure}[h]
		\centering
		\begin{subfigure}{0.2\textwidth}
			
			\includegraphics[width = \textwidth]{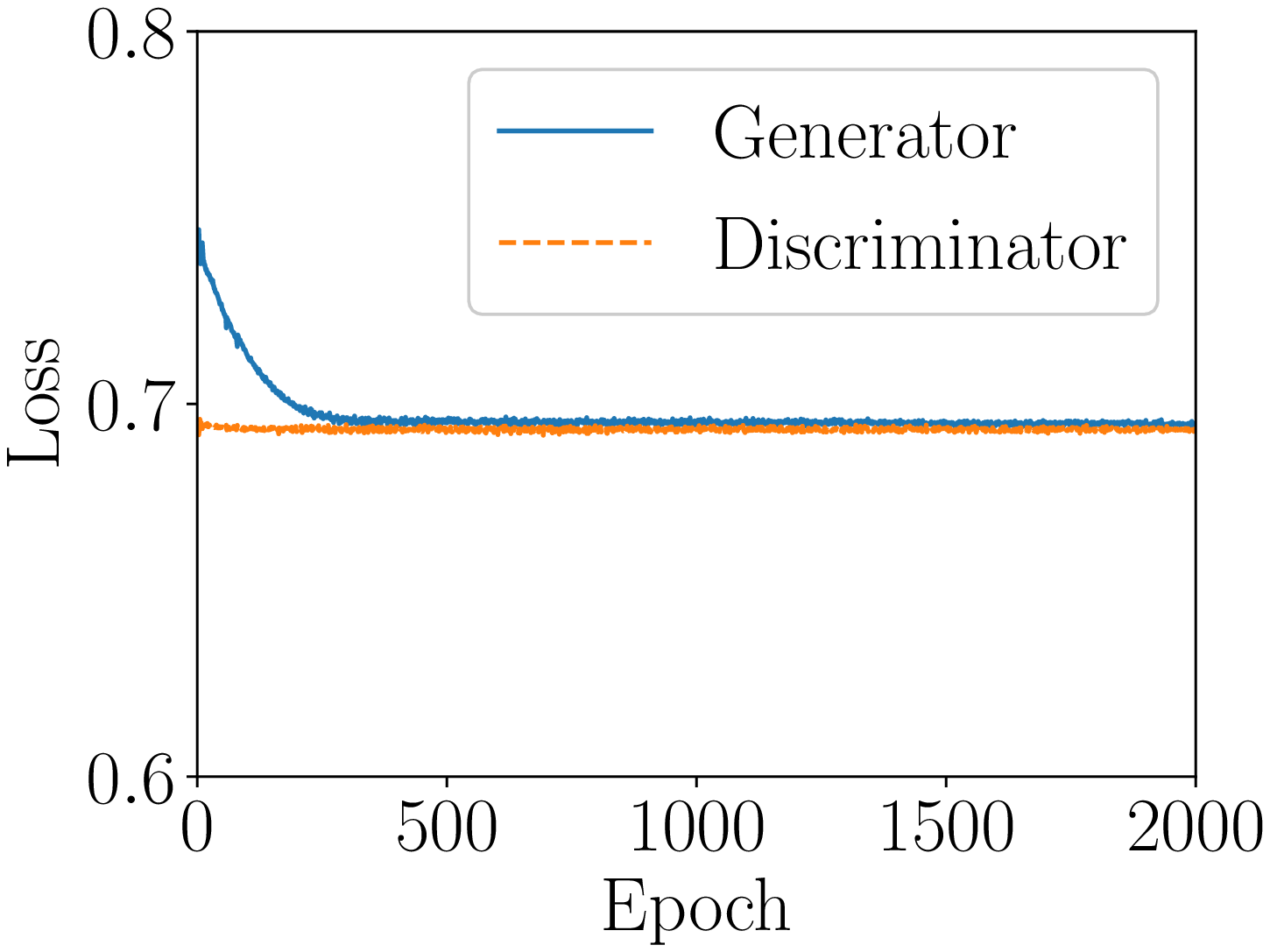}
			\caption{\em}
			\label{fig:Classical_loss}
		\end{subfigure}
		\begin{subfigure}{0.2\textwidth}
			\includegraphics[width = \textwidth]{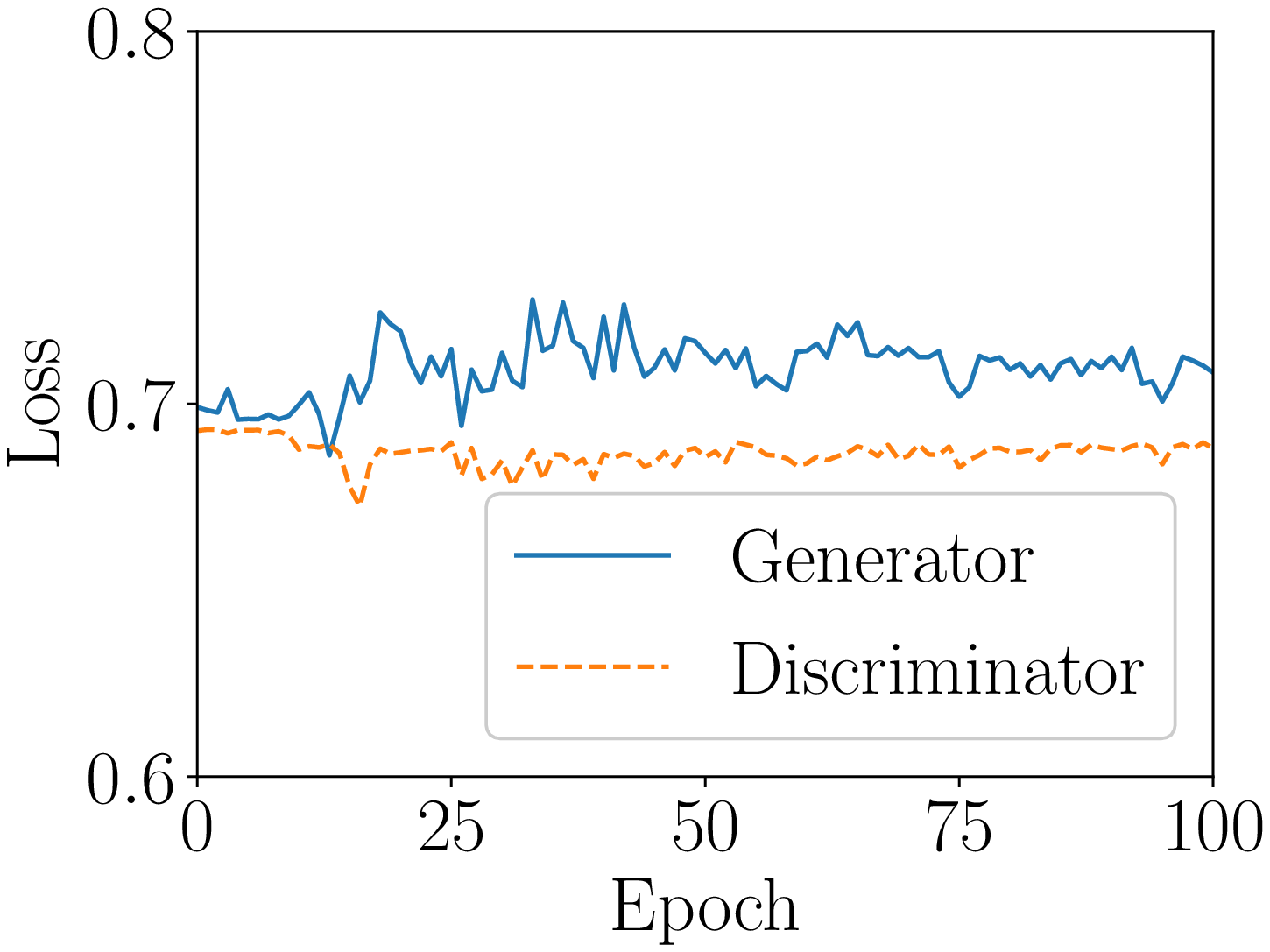}
			\caption{\em}
			\label{fig:CVhybrid_latent3_loss}
		\end{subfigure}
		\begin{subfigure}{0.2\textwidth}
			\includegraphics[width = \textwidth]{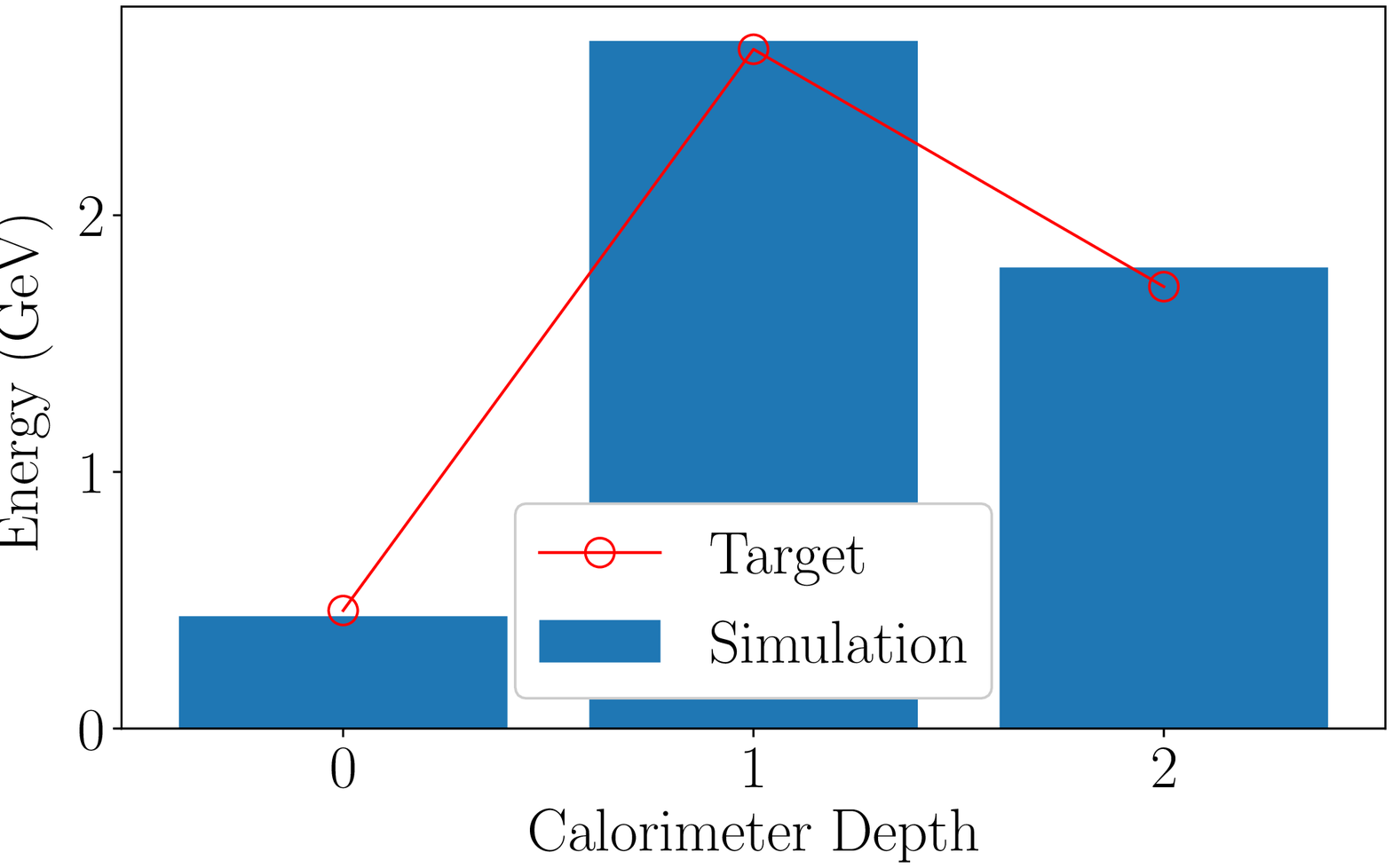}
			\caption{\em}
			\label{fig:Classical_mean}
		\end{subfigure}
		\begin{subfigure}{0.2\textwidth}
			\includegraphics[width = \textwidth]{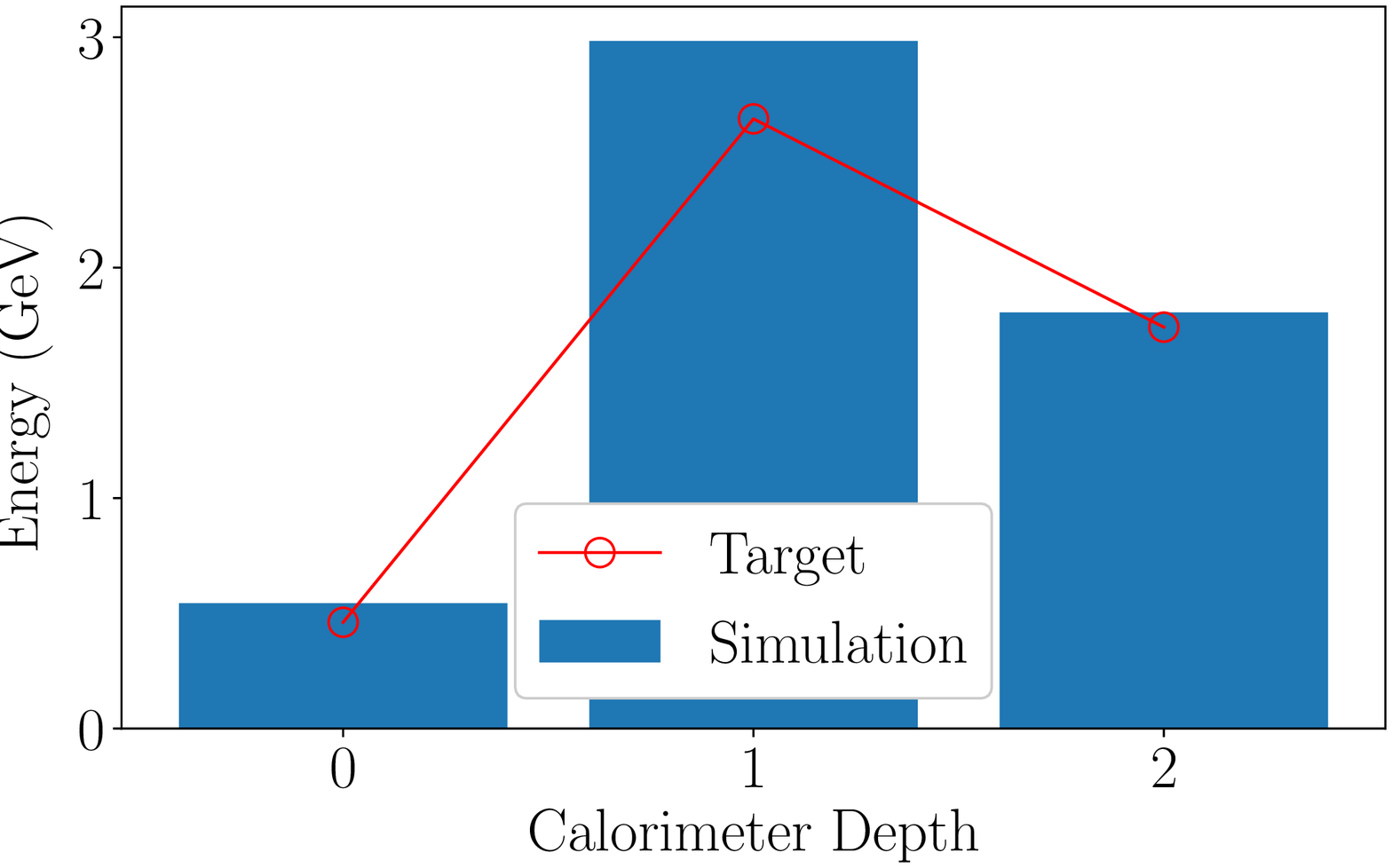}
			\caption{\em}
			\label{fig:CVhybrid_latent3_mean}
		\end{subfigure}
		\begin{subfigure}{0.2\textwidth}
			\includegraphics[width = \textwidth]{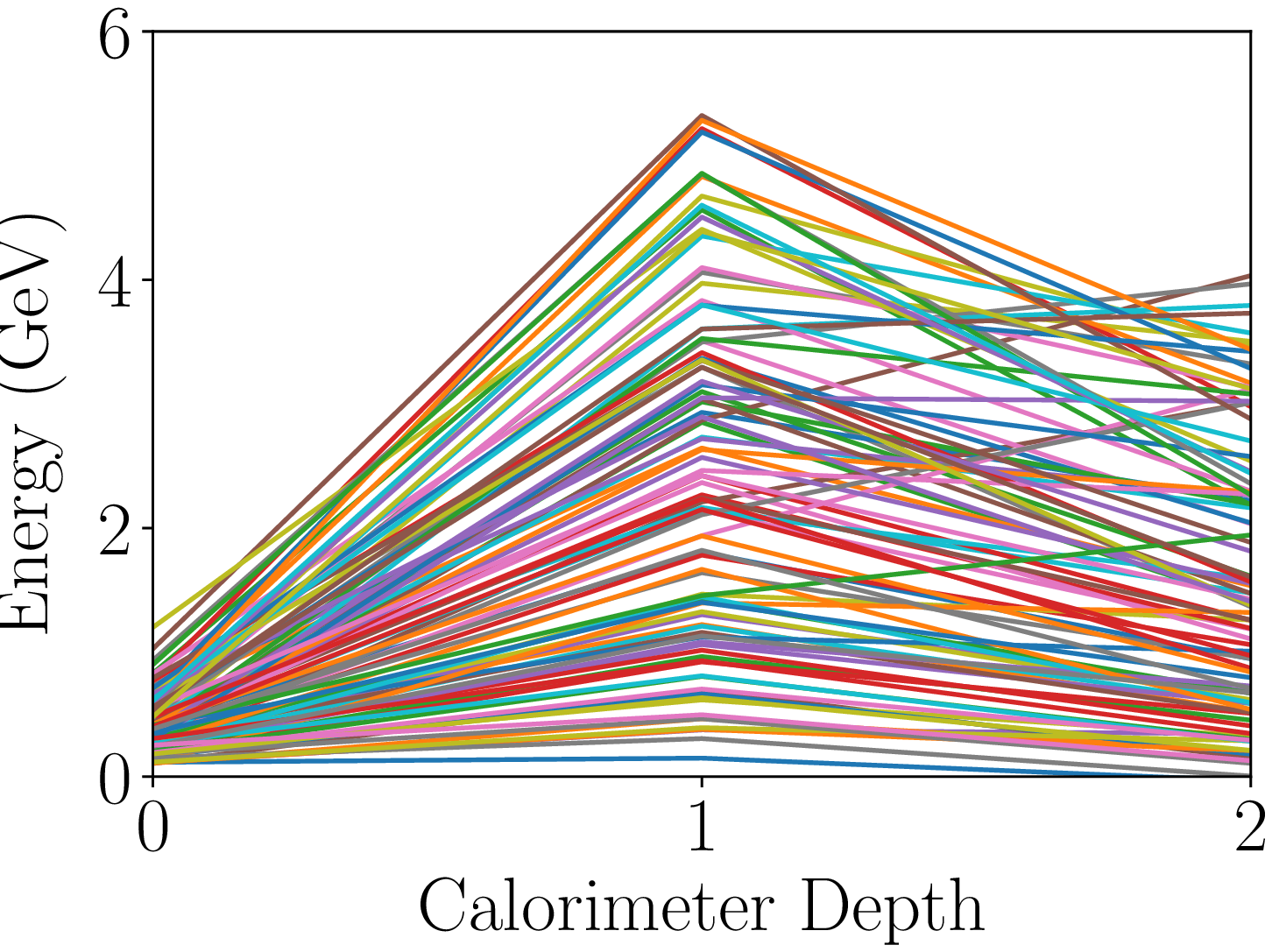}
			\caption{\em}
			\label{fig:Classical_images}
		\end{subfigure}
		\begin{subfigure}{0.2\textwidth}
			\includegraphics[width = \textwidth]{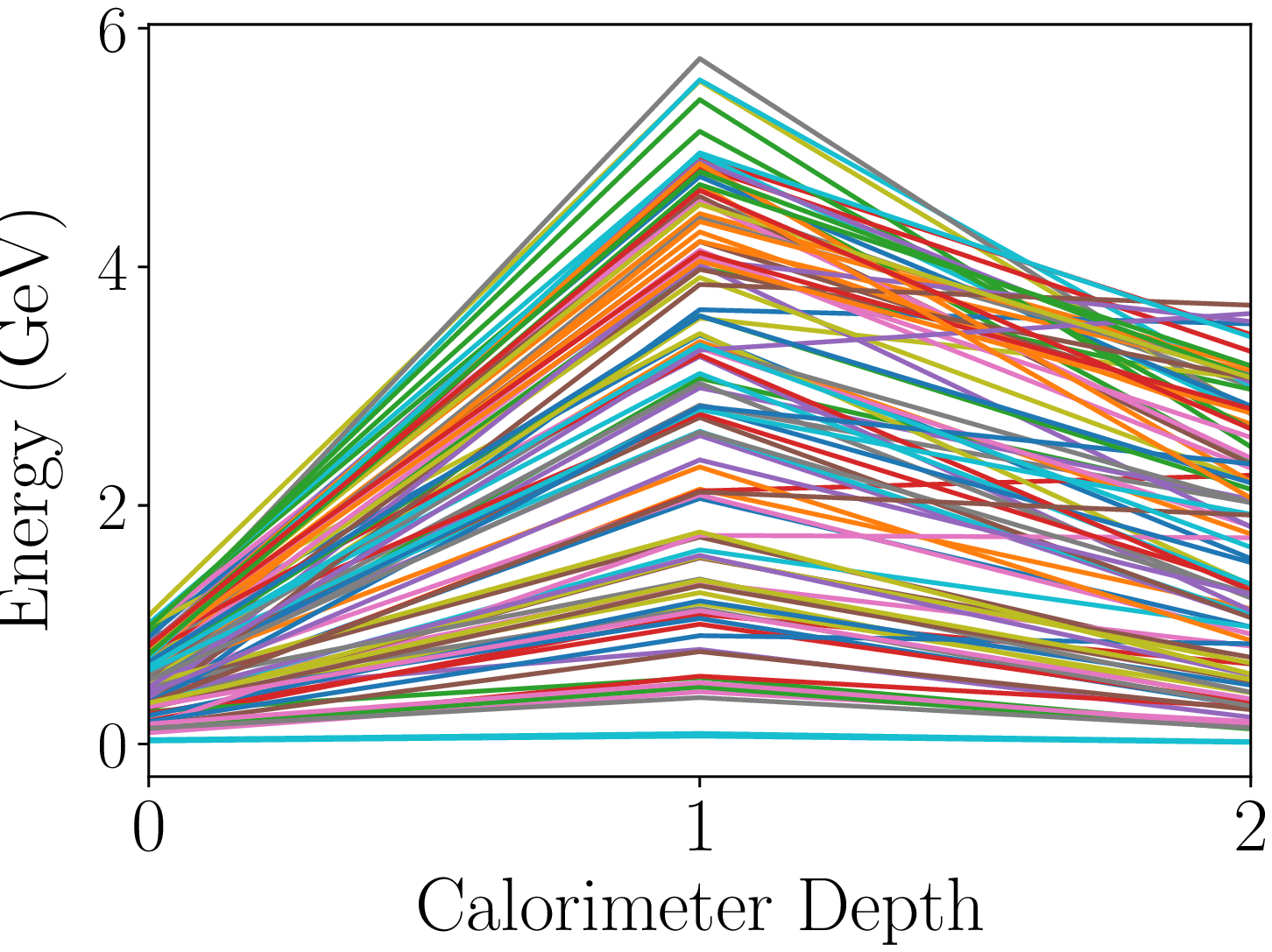}
			\caption{\em}
			\label{fig:CVhybrid_latent3_images}
		\end{subfigure}
    		\caption{\em Losses (a,b), mean images (c,d) and image samples (e,f) obtained from the classical GAN (a,c,e) and the hybrid CV qGAN (b,d,f) simulations with latent space dimension 3. }
		\label{fig:latent3}
	\end{figure}

    \paragraph*{Increasing latent space dimension} To complete this work, hybrid CV qGAN model is tested with latent space dimension 3, i.e., by initializing all three qumodes in the generator with random noises and compared with a classical case as shown on \figref{fig:latent3}. 
    Both models use a classical discriminator with the same structure. But CV qGAN consists of the quantum generator with $d_g=8$, while the classical one employs a classical generator having a similar size as the discriminator. 

    \figref{fig:CVhybrid_latent3_mean} and \figref{fig:CVhybrid_latent3_images} reveal convergence in mean image as well as individual images reproduced by CV qGAN. In addition, unphysical behavior exhibited in the previous simulations is no more observed, as all the samples have positive energy. It is possible to find few samples with a peak at $x = 2$ on \figref{fig:CVhybrid_latent3_images}, which exist in the real samples on  \figref{fig:real_data_3pixels}  Remarkably, CV qGAN can imitate the performance of classical GAN with 264 parameters for the generator, which are 170 times less compared to 44947 parameters used in the classical case. Although it is impossible to make a direct comparison between the models due to large oscillations in loss functions shown on \figref{fig:CVhybrid_latent3_loss}, the study still gives an insight into the size of CV generator recommended for successful training.

	Furthermore, the simulations have demonstrated that CV qGAN reveals convergence before 100 epochs while classical GAN started converging after 1000 epochs, exhibiting an advantage in terms of computational complexity with respect to the time to convergence.
  Unfortunately, the time needed to simulate the quantum generator is still the limiting factor to reach full convergence (350 min per epoch).

	\section{Conclusions and future work}
	\label{sec:conclusion}
    In this work, we suggest a new prototype of qGAN using continuous-variable quantum computing. Thanks to an equivalence between classical and CV neural networks, it is possible to build a CV qGAN, having an architecture similar to the classical GAN. The current CV qGAN model has limitation in generating correctly the targeted image samples and it exhibits a mode collapse failure, which is often observed in the classical GAN. Stability and convergence are improved by increasing the latent space dimension, a result that is consistent with the finding reported in other works concerning the importance of the initialisation step \cite{QGAN_qiskit} . 
    
	To further improve the model, we will explore techniques to speed the training (and in particular the gradient computation step)  while, at the same time increasing the number of qumodes (and therefore the model representational power). The current simulations require a tremendous amount of time (several days) for a small number of qumodes (<5). Different entanglement configuration will be investigated together with mechanisms reproducing regularisation that are usually applied to classical networks, in order to improve the qGAN model and reproduce the real image data accurately in future studies.

\end{document}